%% file: paper.tex
\documentclass{article}
\usepackage{spconf,amsmath,graphicx,url}
\usepackage{mathptmx}
\usepackage{tikz}
\usetikzlibrary{positioning,arrows}

\title{faster-enhancer.c: A Dependency-Free int8 Runtime for
Streaming Speech Enhancement on Commodity CPUs}

\name{Gyeongmin Kim}

\address{Department of Computer Science, Hanyang University, Seoul, Korea\\
kdrkdrkdr@hanyang.ac.kr}

\begin{document}
\ninept
\maketitle

\begin{abstract}
This is an implementation and measurement study of what it costs to run a
streaming speech enhancer on a CPU. We port FastEnhancer-Medium at 48 kHz to
\texttt{faster-enhancer.c}, a C runtime with six int8 GEMM tiers selected at
initialization, leaving architecture and weights untouched. One Apple M2 core reaches 0.069 real-time factor, against 0.230 for
the fp32 ONNX Runtime graph on the same machine, a 3.3$\times$ speedup. A
Galaxy S23+ (Snapdragon 8 Gen 2) reaches 0.096. The speedup comes from specializing every
layer of the runtime around one fixed model. Activation ranges are recomputed
per frame, so no calibration set is needed; the $k=3$ convolutions use Winograd
$F(2,3)$; cross-stage state is fp16; the GRU and the dequantization epilogues
are fused; and nothing is allocated after startup. Over 824 VoiceBank-DEMAND
utterances the engine tracks fp32 to within $-0.006$ PESQ and $-0.08$ dB SNR.
Speed alone does not settle deployment cost. The enhancer holds a fraction of a core
for as long as the microphone is open, so its real-time factor is a duty
cycle. A benchmark races through a file; an audio callback does not. Pacing to
the
6.67 ms deadline costs 4.2$\times$ per frame, saves 49\% of the energy, and
leaves the cheapest core placement missing 96\% of its deadlines. All SIMD
tiers within an architecture family emit byte-identical output. The runtime is
released as a dependency-free library.
\end{abstract}


\section{Introduction}

Neural suppressors have displaced hand-tuned signal processing for real-time
speech enhancement, and the model-side literature has pushed hard on parameters
and multiply-accumulate counts. Deployment did not follow at the same rate. A
model can be small on paper and still be awkward to place in an audio callback,
because shipping it usually means shipping a general inference runtime, keeping
an exported graph synchronized with the training code, and defending
frame-level tail latency rather than average throughput.

Real-time factor is the number that exposes this, read as a duty cycle rather
than a latency. The reference implementation of FastEnhancer
reports RTF 0.1347 for the 48 kHz Medium configuration on an Apple M5
\cite{fastenhancer_repo}. Enhancement is rarely the product. It runs underneath
a call, a recording session, or a transcription pipeline, often for hours, and
usually on hardware weaker than an M5. Work targeting microcontrollers states
the constraint in physical terms \cite{rusci2022tinydenoiser}: a 60 mAh cell
sustains a 20-hour lifetime only near 10 mW average power. Application
processors have a larger budget, not an unlimited one.

This paper takes the FastEnhancer-Medium architecture
\cite{ahn2025fastenhancer} and the 48 kHz weights released in that same
repository; the published paper trains and evaluates at 16 kHz. We change neither architecture nor training, and ask how far
deployment cost falls when every layer of the runtime is specialized around
one fixed streaming model.

Prior work brackets this problem from both sides. RNNoise showed a neural
denoiser can run in real time at 48 kHz on a CPU, with permissively licensed
source \cite{valin2018rnnoise}. Its quality target sits well below current
models; we are trying to get a current model into that shape. DeepFilterNet and DeepFilterNet2
reach 48 kHz real-time enhancement with a native implementation
\cite{schroter2022deepfilternet, schroter2022deepfilternet2}, and GTCRN pushes
the model side to ultra-low complexity \cite{rong2024gtcrn}. On
microcontrollers, TinyLSTMs \cite{fedorov2020tinylstm} prunes and quantizes an
LSTM enhancer, and a mixed FP16-INT8 post-training scheme
\cite{rusci2022tinydenoiser} keeps the recurrent layers at eight bits; both
trade model fidelity for a power budget. General deployment stacks \cite{xnnpack,
kleidiai, gemmlowp, llamacpp} supply kernels but not the model-specific fusion
or the reproducibility contract needed here.

This paper makes four contributions. (1) A training-free int8 port that leaves
architecture and weights untouched: activation ranges are recomputed from each
frame, so there is no calibration set and no sensitivity to one. (2) A
runtime specialized end to end around one fixed model: six int8 GEMM tiers,
Winograd $F(2,3)$, fp16 cross-stage state, fused GRU and dequantization
epilogues, and a state layout that allocates nothing per frame. It reaches
0.069 real-time factor on one Apple M2 core against 0.230 for the fp32 graph,
tracking it to $-0.006$ PESQ over 824 utterances. (3) Measured real-time
factor and joules per second of audio across two silicon tiers and three ISA
tiers, including an inversion of the best tier between devices, made controlled
by byte-identical output across tiers within a family. (4) A deadline-paced
benchmarking protocol, and the finding that racing through a file misstates
per-frame cost by 4.2$\times$ and hides that the cheapest core placement
cannot hold its deadlines. The runtime is released as a dependency-free
library.

\section{Runtime Design}

\subsection{Model contract and streaming state}

\begin{figure}[t]
\centering
\resizebox{0.94\columnwidth}{!}{\input{fig_pipeline.tex}}
\caption{Per-frame runtime graph; shading encodes arithmetic rather than layer
type. Only \texttt{gru\_h} survives the call. Both fp16 buffers are read and
written by the kernels directly, with no pack or unpack pass.}
\label{fig:pipe}
\end{figure}
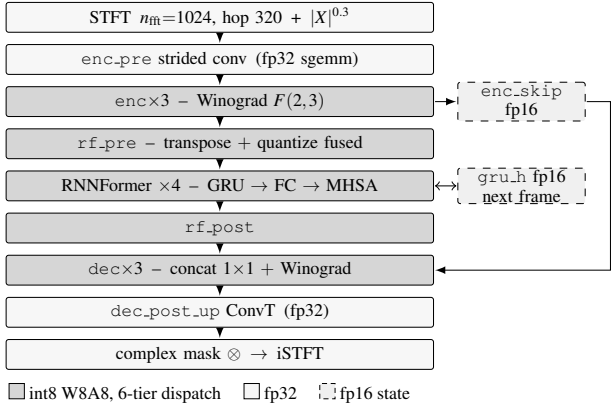

\texttt{faster-enhancer.c} is that specialization; Fig.~\ref{fig:pipe} shows its
per-frame graph. FastEnhancer-Medium predicts a complex mask from a 1024-point STFT, on
magnitudes compressed by $p = 0.3$. It has a strided convolutional
encoder with a $k=8$ stem and three $k=3$ blocks at 96 channels, four blocks
pairing a GRU with four-head attention over 72 channels, and a mirrored
decoder. Each call consumes and produces
320 samples, or 6.67 ms, and the STFT contributes a fixed alignment delay of
704 samples (14.67 ms) with no look-ahead. The public interface is four
functions and one header. All runtime buffers live in a single 432{,}384-byte
state structure allocated once, so nothing is allocated per frame and long runs
cannot drift into allocator jitter. The static library is 162 KiB on macOS
arm64 with no external runtime dependency. The q8 blob is 565{,}108 bytes,
3.7$\times$ smaller than the fp32 ONNX export, and carries 511{,}754
parameters. The runtime is narrow on purpose: CPU only,
single threaded, one global instance, no scalar fallback tier. A host without
the required baseline fails at initialization rather than taking an unverified
slow path.

\subsection{The \boldmath$[-127,127]$ clamp}

Weights are per-output-row symmetric int8 with one fp32 scale per row.
Activations are quantized per frame with a per-tensor asymmetric uint8 range and
stored shifted by $-128$, so signed kernels consume them directly. The range is
recomputed from each frame, so there is no calibration set and no sensitivity
to one. The
zero-point correction folds into the dequantization epilogue through
precomputed row sums, following the integer-only formulation of
\cite{jacob2018quantization}, leaving the hot path as an int8 dot product and
one fused multiply-add.

Both weights and activations are clamped to $[-127,127]$, one code point short
of the full int8 range. Discarding that endpoint buys two things. It makes int16
accumulation provably safe in tiers without a native int8 dot product, since
the worst-case pair sum becomes $2\times127\times127 = 32{,}258$ against a limit
of 32{,}767. It also removes the one value that breaks reproducibility: AVX2
reaches signed products through \texttt{vpsignb}, which cannot negate $-128$.
With the clamp, every tier in an architecture family emits byte-identical
output, at a measured cost of 0.000 PESQ to three decimals.

At first that contract held only inside a single build. Binaries compiled by
Apple clang for M2 and by NDK clang for the Snapdragon part produced different
output for the same input, agreeing on only 22\% of samples. The cause was
neither the kernels, which are integer, nor floating-point contraction, which
we ruled out by rebuilding both binaries with contraction disabled. It was the
analysis window. The engine
built its Hann coefficients at initialization with \texttt{cosf}, which is not
correctly rounded, and the macOS and bionic implementations disagree in the
last bit for hundreds of the 1024 coefficients. The window multiplies every
frame, so one wrong bit separates two builds forever. Computing the window
in double and narrowing once makes the two platforms bit-identical, and costs
nothing at run time because it happens once.

That contract constrains the rest. SSE4.1 was dropped because without FMA3 the
dequantization epilogue becomes a two-step rounding that drifts about one ULP
per operation until hashes diverge, so the x86 floor is AVX2 with FMA3 and
F16C. On AVX-512 an early version used the 14-bit \texttt{rcp14} reciprocal
where the 256-bit path uses 12-bit \texttt{rcp}. The divergence in the GRU
gates surfaced as 44 dB SNR between tiers that were supposed to be identical;
we reproduce the 12-bit estimate by splitting the register into halves. Both
were found by hash comparison, not by listening. No AVX-512 host was
available, so that tier is verified functionally under Intel SDE and carries
no row in Table~\ref{tab:speed}.

\subsection{Kernels and fusion}

Six tiers are compiled into one binary and one is selected at initialization:
ARM NEON, DOTPROD and I8MM, and x86 AVX2, AVX-VNNI and AVX-512 VNNI. Shapes are
fixed, so kernels specialize on the actual $M$, $N$, $K$, and compile-time
alignment assertions let us delete every scalar remainder path.

The binding constraint was rarely the dot instruction; it was how the
activation operand is fed. Load-and-replicate per row held DOTPROD at 62\% of
SDOT peak; pre-packing activations into row quads and indexing lanes reaches
186 GMAC/s, and I8MM needed the same treatment for its row-pair operand. On
AVX-VNNI, broadcasting the activation from memory frees the issue port
\texttt{vpdpbusd} needs; with 12 accumulators the kernel moves from
latency-bound to throughput-bound. AVX2 has
no int8 dot-accumulate and reaches about 24 MAC/cycle, an architectural floor.

Apple M2 is a boundary case worth stating, because it contradicts the obvious
assumption. SMMLA packs twice as many MACs per instruction as SDOT, but M2 issues
SDOT on four pipes and SMMLA on two, so both reach 64 MAC/cycle. I8MM measures
183 GMAC/s, just below the DOTPROD figure above. On this core the I8MM tier is
not why the runtime is fast.

Fusion removes graph-level round trips: transpose folds into quantization,
residuals into GEMM epilogues, and the GRU's two matrices, three gates, hidden
update and fp16 state write are one kernel per tier so int32 accumulators never
reach memory. Softmax consumes int32 scores directly. The $k=3$ convolutions
use Winograd $F(2,3)$ \cite{lavin2016winograd}, and long-lived state is IEEE
binary16, where int8 would lose about 40 dB SNR through the decoder concat.

\section{Evaluation}

\subsection{Protocol}
\label{sec:protocol}

Timing uses two devices: an Apple M2, which provides both FEAT\_I8MM
and FEAT\_DotProd, and a Galaxy S23+ (SM-S916N, Snapdragon 8 Gen 2). The three
x86 tiers build and pass the tier-equality gate, but no x86 host was available
under this protocol, so they are not timed here. Quality is evaluated on the VoiceBank-DEMAND test set
\cite{valentini2017vbdmd} at its native 48 kHz: 824 utterances whose two
speakers and noise conditions the 48 kHz model's training set excludes. It is
also the set the upstream configuration validates on. Timing uses three 48 kHz
RNNoise demo clips at 5 dB, 5569 frames each; percentiles are steady-state,
after the first 200 frames. A separate timing set is sound because the
engine's compute is data-independent -- fixed shapes, no data-dependent
branching, denormals flushed -- and the three clips' p50 agree to within
0.3\%.
The baseline is the exported graph under ONNX Runtime \cite{onnxruntime} in
fp32, on the same machine, timing frame-wise inference only. Package power, cluster frequency and die temperature
are sampled at 5 Hz with \texttt{macmon}, bracketed by idle. Energy is the
integral over the run window minus that idle baseline; the window is located
in the trace from the runner's reported duration, not assumed from a fixed
offset. Each energy cell is three repeats of 180 s of
audio, reported as the median. The runtime and its
regression harness are available at
\url{https://github.com/kdrkdrkdr/faster-enhancer.c} (commit \texttt{7d78dab}).

What each metric can see matters for a fullband model and is usually left
implicit. PESQ is defined by ITU-T P.862 only for rates up to 16 kHz, so it is
computed on a 16 kHz resample and is blind above 8 kHz \cite{itu2005pesq,
torcoli2025pesq}; STOI and eSTOI \cite{taal2011stoi, jensen2016estoi} observe
0--5 kHz, their reference implementation resampling to 10 kHz internally. Waveform SNR,
SI-SDR \cite{leroux2019sisdr}, log-spectral distance at the native rate with a
32 ms window, and SIGMOS \cite{ristea2024sigmos}, a fullband non-intrusive
predictor built on ITU-T P.804, all observe the band a 48 kHz model exists to
reconstruct. The more common DNSMOS \cite{reddy2022dnsmos} is a 16 kHz metric.
Fullband paired test sets remain scarce -- the DNS Challenge fullband tracks
publish no clean references and EARS \cite{richter2024ears} is non-commercial
-- a further reason to evaluate where upstream does. We
report all of them and treat the fullband group as load-bearing; SNR and
SI-SDR agree to 0.001 dB on the port delta, so Table~\ref{tab:quality} lists
SNR alone.

The fp32 reference is generated on the engine's own causal analysis grid
rather than with a default centered STFT, whose frames sit 192 samples away.
That offset is not a multiple of the 320-sample hop and so does not cancel, and
scoring against it would measure framing phase rather than quantization error.
With the grids matched the waveforms align at zero lag.

\subsection{Throughput against baselines}

Table~\ref{tab:speed} gives the device results and the same-machine baselines.
The native runtime reaches 0.069 RTF against 0.230 for the fp32 graph, a
3.3$\times$ speedup, on a model whose largest GEMM is only
$72\times216\times72$ -- shapes at which a general graph executor spends much
of its time on per-operator bookkeeping rather than arithmetic. Per frame the
engine issues 1.60 Mcycle for 46.3 MMAC,
or 28.9 MAC/cycle end to end; against the 64 MAC/cycle SDOT ceiling the frame
runs at 45\% of peak while the GEMM kernels themselves sit at 85--93\%. At the
core's roughly 210 GMAC/s issue peak, 46.3 MMAC puts the RTF floor near 0.033
against the measured 0.069. The gap is STFT, nonlinearities and per-frame
quantization.

The best tier is not the same on every device. On M2 the dispatch-port
symmetry of Sec.~2.3 ties the two tiers to within 1\%, DOTPROD marginally
ahead. On the Snapdragon 8 Gen 2 the
ranking inverts: DOTPROD costs 17.7\% more than I8MM. Each part runs all six
tiers from one binary, and the output hashes agree across tiers and across
both parts, so this is a property of the issue widths rather than of the code.
A build that picked one tier at compile time would be wrong on one of the two.

The S23+ figures are a pessimistic bound. \texttt{adb shell} has no
CAP\_SYS\_NICE, so those runs execute at normal priority, and an application
holding an elevated audio thread would see a tighter tail. The SoC also
hot-unplugs its prime core and refuses a prime-only affinity mask, so every
placement below is a cluster rather than a core.

\begin{table}[t]
\centering
\caption{Median RTF, racing: frames fed as fast as the core accepts them
(Sec.~\ref{sec:duty}). All rows are repeated-median runs under the same
protocol.}
\label{tab:speed}
\begin{tabular}{@{}llccc@{}}
\hline
System & Tier / precision & p50 & p99 & ms/fr.\\
\hline
\multicolumn{5}{l}{\emph{Apple M2, same machine}}\\
ONNX Runtime & fp32 & 0.230 & --- & 1.531\\
\texttt{fe} & I8MM & \textbf{0.069} & 0.084 & 0.458\\
\texttt{fe} & DOTPROD & 0.068 & 0.081 & 0.452\\
\texttt{fe} & NEON & 0.163 & 0.187 & 1.085\\
\hline
\multicolumn{5}{l}{\emph{other devices, \texttt{fe}}}\\
Galaxy S23+ & I8MM & \textbf{0.096} & 0.105 & 0.640\\
Galaxy S23+ & DOTPROD & 0.113 & 0.122 & 0.753\\
Galaxy S23+ & NEON & 0.344 & 0.365 & 2.293\\
\hline
\end{tabular}
\end{table}

\subsection{Quality over the full test set}

Table~\ref{tab:quality} reports the 824-utterance means, grouped by the band
each metric can see. On the full-band metrics the engine tracks the fp32 model
to $-0.079$ dB SNR, $-0.23$ LSD -- the negative LSD meaning the quantized
output is marginally \emph{closer} to the clean target -- and $-0.017$ SIGMOS,
against an enhancement that moves SIGMOS by $+1.064$. Quantization costs 1.6\%
of what the model gains. SIGMOS noise suppression alone is statistically
indistinguishable ($p = 0.90$). On the band-limited metrics the difference is
$-0.006$ PESQ and $-0.0003$ STOI. At $n = 824$ all but the noise term are
significant under a two-sided Wilcoxon signed-rank test ($p < 10^{-4}$). Both
halves of that matter. The port is consistently different from the fp32 model,
and the difference is far below what a listener could hear. Prior
post-training quantization of comparable enhancers reports $-0.06$ PESQ for a
mixed FP16-INT8 scheme and $-0.3$ for uniform int8 \cite{rusci2022tinydenoiser}.

The cost does not grow on harder inputs. Split by input SNR, the PESQ delta is
$-0.003$, $-0.006$, $-0.008$ and $-0.005$ at 2.5, 7.5, 12.5 and 17.5 dB, and
the SNR delta stays within $-0.07$ to $-0.09$ throughout; by noise type it
spans $-0.012$ to $+0.003$. The tier-equality hashes hold on all 824 clips, and
we run them as a gate before any non-trivial change.

\begin{table}[t]
\centering
\caption{VoiceBank-DEMAND test set, 824 utterances at 48 kHz, means; metric
bands as in Sec.~\ref{sec:protocol}. The last row scores the q8 output against
the fp32 output rather than against clean speech.}
\label{tab:quality}
\begin{tabular}{@{}lcc@{\hskip 8pt}ccc@{}}
\hline
& \multicolumn{2}{c}{band-limited} & \multicolumn{3}{c}{full band}\\
Signal & PESQ & STOI & SNR & LSD & SIGMOS\\
\hline
Noisy input & 1.967 & 0.9211 & 8.39 & 14.72 & 2.317\\
fp32 ONNX & 3.060 & 0.9512 & 19.43 & 12.56 & 3.381\\
\texttt{fe} q8 & 3.054 & 0.9509 & 19.35 & 12.33 & 3.363\\
\hline
q8 vs fp32 & 4.610 & 0.9998 & 37.80 & 2.86 & ---\\
\hline
\end{tabular}
\end{table}

\subsection{Duty-cycled operation}
\label{sec:duty}

A benchmark races. It feeds frames as fast as the core will take them, so the
core sits at its maximum clock throughout. A deployed enhancer does not. It
receives one frame every 6.67 ms and idles in between, and the governor
responds by dropping the clock. We measured racing and pacing on the same
audio with the same binary, and on both core types, by requesting
\texttt{QOS\_USER\_INTERACTIVE} for a P-core placement and
\texttt{QOS\_BACKGROUND} for an E-core placement.

Pacing raises per-frame cost by 4.2$\times$ on the P-core, from 0.068 to 0.286
RTF. The engine does identical work in both cases; what changes is the clock it
runs at and the fact that each frame starts after an idle gap. The racing
figure is therefore a floor on what a real audio callback pays per frame.
Both remain real-time. At 0.286 there is still 71\% of the budget left.

\begin{figure}[t]
\centering
\includegraphics[width=0.95\columnwidth]{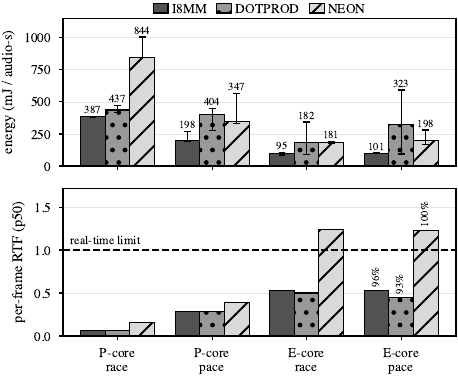}
\caption{Energy per audio-second by core placement and pacing, Apple M2, 180 s
of audio. Bars are medians, whiskers the full range over repeats; the wide
DOTPROD whiskers in the
low-power cells are the measurement resolution, not a property of the tier.
Labels give the deadline-miss rate where it exceeds 1\%.}
\label{fig:duty}
\end{figure}

Racing on the P-core costs 387 mJ per audio-second (Fig.~\ref{fig:duty});
pacing the same tier on the same core costs 198, a 49\% saving for work that
is bit-identical. Race-to-idle is the wrong instinct here. The core finishes
sooner, but it does so at 3.5 GHz, and the voltage that clock costs is never
repaid by the idle that follows. The choice of SIMD tier decides energy as
much as speed.
Falling to the NEON floor costs 844 mJ against 387 racing, and 347 against 198
paced, so the kernel work is worth roughly a factor of two in battery terms on
a device that has FEAT\_DotProd.

The same experiment on the S23+ separates two things macOS conflates: on
macOS, reaching the small cluster means requesting \texttt{QOS\_BACKGROUND},
while Android's \texttt{taskset} sets affinity without touching the scheduling
class. Racing spreads the three placements 5.7$\times$, from 0.090 RTF pinned
to mid-plus-prime to 0.515 on the three little cores. Pacing collapses that to
1.06$\times$: 0.684 on mid-plus-prime against 0.724 on the little cores, which
miss 0.90\% of deadlines against 0.05\%. A racing benchmark says the little
cluster is
unusable; at deployment duty it is within 6\% of the big one, because near
10\% duty the governor holds every cluster at its floor. I8MM holds the
deadline on all three placements; NEON, tried on the little cluster, misses
every one. Energy could not
be measured there. Without root there is no power-stats HAL or readable current
node, and the model-based estimator cannot see pacing.

Back on M2 the E-core columns are cheaper still, 95 mJ racing and 101 mJ
paced. That saving is unreachable through the mechanism that produced it.
\texttt{QOS\_BACKGROUND} lands on the E-cluster and brings timer coalescing
with it. The paced E-core misses 96\% of its deadlines even though the median
frame takes 3.5 ms of a 6.67 ms budget. The compute fits; the wake-up does
not. The deployable point here is the paced P-core at 198 mJ per
audio-second and 0.05\% missed deadlines.

\subsection{Sustained operation}

A 37-second benchmark does not test the claim that motivated the work. We paced
both devices for 30 minutes, 270{,}000 frames, on the tier each would deploy.

M2 drifts but stays inside the budget. p50 rises to 0.2987 RTF from 0.2863 over
180 s, p99 is 0.5648, and 0.124\% of frames miss their deadline against 0.05\%
short. Heat is not the mechanism. Die temperature holds near 36$^{\circ}$C, and
the P-cluster opens at 1102 MHz then settles at 784, 824 and 833 over the last
three quarters, which is a governor hunting rather than a part throttling.

The phone crosses the budget. The same 30 minutes gives p50 0.7215 but p99
1.0139, and misses 1.77\% of deadlines against 0.90\% over 5569 frames, with
per-frame cost rising 35\% from the first hundred frames to steady state. Its
die temperature is flat at 38.2$^{\circ}$C and cluster frequency wanders
without trend, averaging 895, 953, 1062 and 995 MHz by quarter. Whatever widens
the tail is scheduler behaviour, which is the more awkward answer because heat
would at least be predictable. That rate is an upper bound for the same reason
the throughput figures are: at normal priority a late wakeup counts as a slow
frame. Both devices degrade from the short run to the long one, and only one of
them degrades past the deadline, so a phone deployment has to be validated at
deployment duration.

\subsection{Ablations}

Profiling puts the largest per-frame bucket on the fused GRU kernel at 82.2 us,
then QKV projection at 36.3 us, with attention and Winograd blocks at 22--26 us
each; the buckets nest, so they do not sum to the frame. That map resolves an
apparent inconsistency. Winograd cuts the multiply footprint 13\%, from 53.4
to 46.3 MMAC per frame, but removing it raises p50 by 30.1\%, because the
direct path also pays im2col traffic and a layout change the transform avoids.
Multiplies are not the currency.

Four ablations, each a median over the same protocol and each stated against
its own group's baseline. Falling to the NEON floor costs $+138.5$\% of p50
(0.0685 to 0.1634), which is what a device without FEAT\_DotProd pays; forcing
DOTPROD on M2 changes p50 by $-1.0$\%, i.e. nothing. Replacing the fast SIMD
transcendentals with the accurate polynomial build costs $+20.1$\% (0.0682 to
0.0819), and removing Winograd the $+30.1$\% above (0.0692 to 0.0900).
Link-time optimization is neutral at $-0.6$\% and stays opt-in. The
transcendental and Winograd ablations change the output bits, so they are
speed measurements,
not quality-neutral substitutions; the accurate-polynomial build is the more
faithful path, so its 20.1\% is the price of an accuracy axis we did not
measure.

\section{Discussion and Limitations}

The deployment gap closes mostly without touching the model, and no single
technique dominates: fixed shapes, integer arithmetic, one-time dispatch, a
state layout that allocates and converts nothing per frame, and fusion where
the scale computation allows. Where we hit a wall it was architectural: the
AVX2 int8 floor, the M2 dispatch-port symmetry. And bit-identity does not
cross architecture families, only compilers and devices within one.

Three limits apply. The quality figures use a test set that upstream also
validates on, so they measure the port faithfully but not the model. The x86
tiers are verified for bit-equality but not timed. And some fusions rest on
profile attribution rather than one-switch-off ablations.

What the result buys is a deployable artifact rather than a benchmark number:
a published enhancer, retrained on nothing, running at a fraction of one core
on hardware people already own, at a quality its own fp32 graph would not be
distinguished from. We release it as a dependency-free library so that the
next system does not have to rebuild the runtime to find out.

\bibliographystyle{IEEEbib}
\bibliography{refs}

\end{document}

%% file: fig_pipeline.tex
\begin{tikzpicture}[
  font=\scriptsize,
  node distance=1.55mm,
  box/.style   = {draw, rounded corners=1pt, minimum height=4.1mm,
                  text width=58mm, align=center, inner sep=1pt},
  i8/.style    = {box, fill=black!16},
  f32/.style   = {box, fill=black!3},
  st/.style    = {draw, dashed, rounded corners=1pt, fill=black!6,
                  minimum height=3.6mm, text width=17mm, align=center,
                  inner sep=0.7pt},
  ar/.style    = {-latex, thin, shorten >=0.4pt, shorten <=0.4pt}
]

\node[f32] (stft)  {STFT \,$n_{\mathrm{fft}}{=}1024$, hop 320 \,+\, $|X|^{0.3}$};
\node[f32,below=of stft]   (encpre) {\texttt{enc\_pre} strided conv \,(fp32 sgemm)};
\node[i8, below=of encpre] (enc)    {\texttt{enc}$\times$3 \,--\, Winograd $F(2,3)$};
\node[i8, below=of enc]    (rfpre)  {\texttt{rf\_pre} \,--\, transpose $+$ quantize fused};
\node[i8, below=of rfpre]  (rf)     {RNNFormer $\times 4$ \,--\, GRU $\to$ FC $\to$ MHSA};
\node[i8, below=of rf]     (rfpost) {\texttt{rf\_post}};
\node[i8, below=of rfpost] (dec)    {\texttt{dec}$\times$3 \,--\, concat $1{\times}1$ $+$ Winograd};
\node[f32,below=of dec]    (decpost){\texttt{dec\_post\_up} ConvT \,(fp32)};
\node[f32,below=of decpost](mask)   {complex mask $\otimes$ \,$\to$\, iSTFT};

\foreach \a/\b in {stft/encpre, encpre/enc, enc/rfpre, rfpre/rf, rf/rfpost,
                   rfpost/dec, dec/decpost, decpost/mask}
  \draw[ar] (\a) -- (\b);

\node[st, right=3.2mm of enc]  (skip) {\texttt{enc\_skip}\\fp16};
\node[st, right=3.2mm of rf]   (gruh) {\texttt{gru\_h} fp16\\next frame};
\draw[ar] (enc.east)  -- (skip.west);
\draw[ar,<->] (rf.east) -- (gruh.west);
\draw[ar] (skip.east) -- ++(3.4mm,0) |- (dec.east);

\node[anchor=north west, align=left, inner sep=1pt]
  at ([yshift=-1.4mm]mask.south west)
  {\tikz\draw[draw=black,line width=0.3pt,fill=black!16]
      (0,0) rectangle (2.1mm,2.1mm); int8 W8A8, 6-tier dispatch \quad
   \tikz\draw[draw=black,line width=0.3pt,fill=black!3]
      (0,0) rectangle (2.1mm,2.1mm); fp32 \quad
   \tikz\draw[draw=black,line width=0.3pt,dash pattern=on 0.7mm off 0.5mm,
              fill=black!6] (0,0) rectangle (2.1mm,2.1mm); fp16 state};

\end{tikzpicture}